**Worst Case Resistance Testing:**

**A Nonresponse Bias Solution for Today's Behavioral Research Realities**

Stephen L. France

Associate Professor of Quantitative Analysis

Mississippi State University

PO Box 9582, Mississippi State, MS 39762

Email: sfrance@business.msstate.edu

Frank G. Adams

Associate Professor of Marketing

Mississippi State University

PO Box 9582, Mississippi State, MS 39762

Email: fadams@business.msstate.edu

V. Myles Landers

Assistant Professor of Marketing

Mississippi State University

PO Box 9582, Mississippi State, MS 39762

Email: vlanders@business.msstate.edu

**Worst-Case Resistance Testing:**

**A Nonresponse Bias Solution for Today's Behavioral Research Realities**

**Abstract**

This study proposes a method of nonresponse assessment based on meta-analytical file-drawer techniques, also known as worst-case resistance testing (WCRT), and suitable for a wide range of data collection scenarios. A general method is devised to estimate the number of significantly different nonrespondents it would take to significantly alter the results of an analysis. Estimates of nonrespondents can be plotted against effect sizes using "n-curves", with similar interpretation to p-curves or power curves. Variants of the general method are derived for tests of means and correlations. A sample using a well-established survey instrument from previous behavioral research is used to test the method. The results suggest that employing worst-case resistance testing can be used on its own or in conjunction with wave analysis to precisely flag nonresponse risks.

# Introduction

All quantitative empirical methods rely on the assumption that the sample participants represent the population of interest sufficiently to justify extrapolation of findings beyond the sample measured (Chesney & Obrecht, 2012). However, some portion of the participants solicited in almost every study do not respond, and as the proportion of those non-respondents grows larger, the study's results suffer from potential bias (Boyd & Westfall, 1965). This participation nonresponse bias is the focus of this paper.

Participant nonresponse bias has been attributed to the variation of characteristics between respondents and nonrespondents (Deming, 1953), and this variance has the potential to confound the variance observed between the constructs measured in any given empirical test (Groves and Peytcheva, 2008) and introduce bias into statistical tests. This bias can be thought of as a type of selection bias and unlike bias for nonresponse of individual items this bias cannot be corrected without gathering data from nonrespondents (Berg, 2005).

The growing use of internet surveys for behavioral survey research has changed the nature of survey response and nonresponse. While scholars have well accepted means of assessing nonresponse bias – most notably, wave analysis – those methods were developed based upon physical mail collection of surveys. By contrast, much survey and experimental research today employs electronically curated samples that can be gathered in hours, or even minutes and that does not have well defined participant response data.

Accordingly, this study develops a set of methods independent of survey delivery mode allowing researchers to examine the robustness of statistical tests against participant nonresponse bias[1] by calculating the number of cases needed to reverse a statistical test over a range of different effect sizes. The resulting "n-curves" provide similar insight to similar methods such as power curves and p-curves and provide a measure of robustness for the results of statistical tests in situations where participant nonresponse may affect the results and conclusions from such tests. This study then tests the proposed methods on an empirical survey of customer satisfaction and shows how these methods can be used on their own or combined with wave analysis to flag statistical tests where nonresponse bias may be problematic. All data files and code used in the creation of this paper are available at https://github.com/MDSOPT/WCRT.

## Background

In recent years, dedicated efforts have examined practices such as "p-hacking" to recall the academy to replicable research methods (Simmons et. al, 2011). Similarly, recent failures to replicate psychological research have been a cause for concern (Stanley et al., 2018), leading to a call for more transparency in research (Inman et al., 2018). This includes reporting data collection techniques, power analysis, effect sizes and potential biases that might influence results.

Arising from a combination of sampling error and coverage error, nonresponse error results in a sufficient difference between the data sought by a researcher and the data actually obtained to compromise a study's validity (Collier & Bienstock, 2007). Conceptually, nonresponse error holds that the potential responses of subjects who do not answer a solicitation to participate in a given research study might be different enough from the responses recorded to alter the findings of the study and higher nonresponse rates can negatively affect the representativeness of a sample (Cook et al., 2000). When records of response data are available, it is quite easy to assess participant nonresponse, but "there is no magical response rate below which an observed mean, standard deviation, or correlation becomes automatically invalid" (Newman 2009, pp. 7). Still, the larger the percentage of the solicited sample measured, the lower the error resulting from nonresponse bias tends to be (Olson, 2006).

---

[1] For the sake of parsimony, as this paper focuses on participant nonresponse issues, subsequent mentions of nonresponse refer to participant nonresponse rather than item nonresponse.

Scholars have developed several procedures to adjust survey results to account for survey nonresponse bias, as detailed in Table 1 (following Halbesleben & Whitman, 2013), but they all inherently rest on a key assumption: "…respondents and nonrespondents within a weighting class have the same values on key variables…" (Groves 2006, pp. 653). Accordingly, attempts to assess nonresponse bias rely on an assumption that differences causing some solicited subjects to forego answering a survey are related to how a nonrespondent might react to a study's constructs of interest. Based on this assumption, for decades, the most widely used method of assessing nonresponse bias has been Armstrong and Overton's wave analysis technique (1977).

Table 1: Summary of Nonresponse Bias Assessments and Remedies

| Technique | Description | References |
|---|---|---|
| Comparison of Sample and Population | Compare demographics of known population characteristics to collected sample characteristics. | Armstrong and Overton (1977); Groves (2006); Beebe, et al. (2011) |
| Wave Analysis | Compare early respondent answers to later respondent answers. | Armstrong and Overton (1977) |
| Follow-up Analysis | Obtain responses from subjects who did not respond to the original data collection in order to test for differences. | Aiken (1981), Sosdian and Sharp (1980) |
| Bayesian Analysis | Utilize Bayes rule to estimate nonresponse data, assuming independence of attributes and | Daniel and Schott (1982) |

| | known characteristics across respondents and nonrespondents | |
|---|---|---|
| Passive and Active Nonresponse Analysis | Attempt to assess why active nonrespondents declined to participate through focus groups, interviews, and surveys about the original data collection. Resend the survey to address passive nonrespondents. | Rogelberg and Stanton (2007); Rogelberg, et al. (2003); Roth (1994) |
| Interest-level Analysis | Include questions about subject interest in the survey topic among the measured items and statistically control for interest when analyzing responses. | Rogelberg and Stanton (2007); Rogelberg, et al. (2000) |
| Benchmarking | Compare sample demographics with those of other studies of similar phenomena to see if there are inconsistencies of means or standard deviations. | Rogelberg and Staton (2007) |
| Replication | Conduct multiple surveys using different samples to assess whether findings remain consistent. | Rogelberg and Staton (2007) |

**Wave Analysis**

Simply put, wave analysis compares relationships between variables observed among early respondents to a measurement instrument with those observed among later respondents (Armstrong & Overton, 1977). "The basic assumption … is that subjects who respond less readily are more like those who do not respond at all than those who do respond readily (i.e., those who respond sooner and those who need less prodding to answer)" (Kanuk & Berenson, 1975, pg. 449). The method poses that a lack of significant difference between early and late respondents to a research solicitation implies that potential subjects that did not respond do not represent observations that might alter an analysis's results.

For all its long-proven utility (over 19,000 citations at this writing), wave analysis was explicitly built around mail surveys, which generally required considerable periods of time to collect (Kanuk & Berenson, 1975) and where information on early and late response waves can easily be found. Studies employing postal mail and citing wave analysis have including follow up prompts of up to four weeks (Diamantopoulous & Winklhofer, 2001; Mohr & Spekman, 1994; Sirdeshmukh et al. 2002). Even surveys distributed over email have noted time spent awaiting response from subjects (Pavlou 2003). As of 2007, internet surveys constituted the majority of surveys, and as of 2020 the vast majority of surveys are completed via the internet (Daikeler et al., 2020). The "internet age" of surveys has seen a growth of third-party survey platforms, such as Qualtrics and Prolific, who recruit participants well in advance of any study, and pay participants fees to complete studies. The resulting participants are more likely to reply quickly because they have pre-agreed to participate in studies (Qualtrics, 2020). On some research platforms, such as Prolific (Peer et al., 2017) and the Amazon Mechanical Turk (Chandler et al., 2019), potential respondents when logging on, will pick from a list of potential surveys or work to complete. In this situation, unless user click/screen viewing behavior is analyzed, it is difficult to identify and quantify nonresponses (e.g., Boas et al. 2020; Paolacci et al. 2010) and the early and late response waves required by wave analysis.

**Resampling**

A typical means of addressing potential nonresponse bias is to simply resolicit sample nonrespondents (Aiken, 1981; Hartman, et al. 1986; MacDonald, et al. 2009), often employing shorter surveys that assess only the items whose constructs are of critical importance to the observed findings, to look for differences from the findings of the original (Lambert & Harrington, 1990). However, the absence of a specific response rate below which nonresponse bias is considered problematic (Newman, 2009) implies that supplemental sampling – whether among the originally solicited group, or from a different group of potential respondents – may not necessarily address nonresponse bias of a sample relative to the population of interest. A different potential solution may lie in meta-analysis techniques to address a bias issue known as the file drawer problem.

**Meta-Analysis and the File Draw Problem**

The file drawer problem is a term used in meta-analytic literature to describe a conceptual, but quantifiable sampling bias. Because meta-analyses examine the standardized results of extant literature, they are presumed to be biased by the tendency of statistically significant findings to achieve academic publication, and the corollary tendency of non-significant results of similar phenomena never entering the scholarly body of knowledge (Rosenberg, 2005). The direst assumptions hold that 95% of contrary findings do not survive the academic publication process, and that the body of knowledge is, therefore, a victim of Type 1 error (Rosenthal, 1979).

Rosenthal proposed a solution to the file drawer problem, sometimes known as worst-case resistance testing (1979), or fail-safe number calculation (Rosenberg, 2005). The technique calculates the number of studies required to significantly alter an observed mean of effect sizes, assuming the hypothetical unobserved studies have a collective mean significantly different than that of the observed effect sizes. As this calculated number of studies increases, the likelihood of a file drawer bias decreases. In other words, the larger the effect size observed in tests of a given sample, and/or the less stringent the standard of testing significance, the more hypothetical contradictory cases it would take to cast doubt on the observed findings. The technique has been used in varying meta-analytic studies including (but by no means limited to) electronic word of mouth (Babić et al. ,2016), interstitial space impacts on consumer appeal (Sevilla and Townsend 2016), and consumer responses to humanoid robots (Mende et al., 2020).

The nonresponse bias problem is very similar to the file drawer problem in that both seek to assess a difficult-to-quantify bias of findings stemming from uncollected data presumed to contradict results based on observed data. It stands to reason that the file drawer solution – WCRT – should also be efficacious in assessing nonresponse bias.

## Methodology

To illustrate how file-draw concepts can be applied to the nonresponse bias problem, the problem is given in general in terms of the basic NHST (null hypothesis significance test) paradigm. Though this paradigm has been much criticized (e.g., Gill, 1999; Hubbard & Armstrong, 2006; Hunter, 1997; Schneider, 2015) it is still by far the predominantly used framework for building theory in empirical management and social science research.

Furthermore, most alternative approaches proposed to replace NHST also have criticisms, for example the use of confidence intervals for inference leads to the same "inverse inference" that is criticized in NHST testing and Bayesian analysis requires specification of prior distributions, which can be conceptually difficult (e.g., Masson, 2011; Trafimow, 2017). While at least one journal has banned significance testing (Woolston, 2015), most journals and scientific associations in the behavior sciences and business disciplines have focused on best practice to improve the use of NHST results and to put these results into context.

Scholars have advanced several recommendations to improve the implementation of NHST methods. These include putting p-values into context and avoiding erroneous overly strong conclusions from p-values (Wasserstein & Lazar, 2016), focusing on the magnitude and size of any statistical effect and incorporating information from prior beliefs (Harvey, 2017), reporting of descriptive statistics and reporting guidelines for major statistical tests (JCR, 2021), including detailed graphs and discussions of effects and utilizing robust error statistics (Schwab et al., 2011), and calculating power values for each statistical test and ensuring that the Type II error rate ($\beta$) is less than 0.05 when making conclusions on a lack of "effect" relative to a null hypothesis (Baroudi & Orlikowski, 1989; Cashen & Geiger, 2004).

A theme in most of the rules and suggestions described above is the "triangulation" of NHST results with other metrics to build evidence for hypothesis test conclusions. As such, the methodology described in this paper fits in with this theme. The aim is to provide a set of measures of robustness of statistical results to problems caused by nonresponse bias. However, the methods described can be used beyond the realm of nonresponse bias to examine robustness to other sources of error, such as the experimental design.

In this study, the WCRT methodology is described using a generic NHST hypothesis testing procedure. Examples are included for problems with simple hypothesis testing of means and of correlations, where equations are given for "finding the number of additional studies" required to negate a conclusion and then models are developed to solve these equations.

**The General Model**

The general problem is outlined as follows: Consider a situation with a NHST performed on data collected from a survey. The purpose of the test is to find sufficient evidence to reject a null hypothesis ($H_0$) in favor of an alternative hypothesis ($H_A$). There is some critical value at which enough evidence is gathered so that the researcher flips from failing to reject the null hypothesis to rejecting the null hypothesis. If the researcher finds enough evidence to reject $H_0$, but H₀ is in fact true, then the researcher is considered to have committed a Type I error, with a probability denoted as α. The value of α is usually defined in terms of extreme results in the distribution of expected sample values in the H₀ distribution, which can be denoted as $\alpha = P(R|H_0)$, where R is rejection of H₀. Given the distribution of H₀, H₀ is rejected if there is enough evidence, operationalized by the sample statistic being far enough away from a "null effect" in a sampling distribution.

A researcher will often make the "opposite assertion", that given insufficient evidence to reject H₀, one can conclude that H₀ is in fact true. However, there is a danger with this assertion in that researchers may assume a trivial effect without understanding the implications of the power of the statistical test (Baroudi & Orlikowski, 1999; Cashen & Geiger, 2004; Sawyer & Ball, 1981). If the researcher fails to reject H₀ and in fact H$_A$ is true, then the researcher has made a Type II error, i.e., $\beta = P(R^c|H_A)$, where the power of the test is $1 - \beta$. An issue here is that H$_A$ can take multiple values and that the power varies with the "effect size" difference between H₀ and the H$_A$ used to calculate power. Solutions to this issue include calculating power using a reasonable effect size based on prior studies, standard small, medium, and large effect sizes (Cohen, 1992), and graphing power values across a range of effect sizes, a "so called" power curve (Faul et al., 2007).

In the context of nonresponse bias and WCRT, the focus is to find the number of non-respondents who can reverse a statistical conclusion and use this as a measure of robustness of the solution. But how is the effect size for these studies chosen? Is it a "zero effect", the opposite effect, or a smaller effect in the same direction? The methodology outlined in this paper mirrors the work described above in choosing effect sizes for power analyses. The number of non-respondents needed to reverse a statistical test can be calculated for a range of feasible effect sizes, which can be estimated from wave analysis or by examining effect sizes for similar studies. These values can be plotted, creating an "n curve", which is similar to curves used for determining quality bounds for confidence intervals (e.g., Trafimow, 2018) or p-curves used to map sample sizes for different p-values at different power levels (Simonsohn et al., 2014).

At the core of the analyses in this paper is the idea of a standardized effect size (Cohen 1998). An effect size can be thought of as a quantitative measure of the phenomenon being studied (Kelley & Preacher, 2012). For example, for a single sample t test, the effect size *d*, is given in (1).

$$d = \frac{\bar{x} - \mu_0}{s} = \frac{(\bar{x} - \mu_0)/\sqrt{n}}{s/\sqrt{n}} = \frac{t}{\sqrt{n}} \tag{1}$$

, where $\bar{x}$ is the sample mean, $s$ is the sample standard deviation, $\mu_0$ is the hypothesized population mean, and *n* is the sample size. This invariance towards *n* is particularly useful for large sample size experiments, as effect sizes can put into context results that are significant with only a small effect size, but a very large sample size (e.g., Coe, 2002). Different statistical tests have different effect size calculations. For example, effect sizes for the comparison of two group means, such as Cohen's d and Glass's g, have a similar format to the effect size given in (1), while for Pearson's correlation, the sample regression coefficient *r* is often used as a measure of effect size (Hemphill, 2003).

In the context of a WCRT analysis of a NHST test, we define a general effect size $\phi$, which can be substituted by the appropriate metric for a specific test (e.g., *d* for sample mean tests). Consider the following situations:

1. With a sample size of $n_1$ there is enough evidence to reject $H_0$. There is an effect size $\phi_1$ that is associated with the test. We wish to find $n_2$, where this is the number of items or non-respondents with effect size $\phi_2$ required to negate the result, so that H₀ is no longer rejected.

2. As above, but with a sample size of $n_1$ there is not enough evidence to reject H₀ and a sufficient $n_2$ with effect size $\phi_2$ required to negate the result, so that H₀ is now longer rejected.

A set of candidate effect sizes needs to be defined for $\phi_2$. This is key to the methods in this paper and the appropriate range can be informed by previous research, the results from a wave analysis of the data, and the effect size $\phi_1$ (for example, if there is a significant effect, an effect size greater or equal to $\phi_1$ and in the same direction is not going to negate the hypothesis test). For each $\phi_2$, the procedure will give the $n_2$ value needed to reverse the result of the statistical test.

**Inference for Single Sample t-test**

Consider a single sample t-test of a population mean being equal to hypothesized mean $\mu_0$. The notation is as per (1) and the null hypothesis is $H_o: \mu = \mu_0$. The methodology outlined in this section covers both two tailed tests where the alternative hypothesis is defined as $H_a: \mu \neq \mu_0$ and one-tailed tests where the alternative hypothesis can be defined as $H_a: \mu > \mu_0$ or $H_a: \mu < \mu_0$. The test statistic derived from the sampling distribution is defined as (2) by rearranging (1).

$$t = \frac{(\bar{x}_1 - \mu_0)}{\frac{s_1}{\sqrt{n_1}}} = d_1\sqrt{n_1} \qquad (2)$$

Here, the subscript 1 indicates that the sample values are based on the responses, while the subscript 2 will be used for the sample values for hypothesized nonresponses. The t distribution has $n$-1 degrees of freedom and varies with $n$. Let $t^*$ be the critical boundary between rejecting and failing to reject H₀. Dependent on $n$ and the strictness of the test (using the Type I error α), H₀ is rejected if $|t| > t^* = t_{\alpha/2}$ for a two tailed test and $t > t^* = t_\alpha$. Here we consider four different scenarios[2].

1. For a one-tailed upper test or a two-tailed test with $t > 0$, H₀ is rejected as $t > t^*$. We wish to find, for some **nonrespondents** with effect size $d_2$, the $n_2$ for required to reverse this conclusion, so that $t \leq t^*$.

2. For a one-tailed upper test or a two-tailed test with $t > 0$, H₀ is not rejected as $t \leq t^*$. We wish to find, for some **nonrespondents** with effect size $d_2$, the $n_2$ required to reverse this conclusion, so that $t > t^*$.

3. For a one-tailed lower test or a two-tailed test with $t < 0$, H₀ is rejected as $t < t^*$. We wish to find, for some **nonrespondents** with effect size $d_2$, the $n_2$ required to reverse this conclusion, so that $t \geq t^*$.

4. For a one-tailed lower test or a two-tailed test with $t < 0$, H₀ is not rejected as $t \geq t^*$. We wish to find, for some **nonrespondents** with effect size $d_2$, the $n_2$ required to reverse this conclusion, so that $t < t^*$.

A simplifying assumption is to assume that $s_1 = s_2$, i.e., the nonresponse and response data have the same standard deviations. However, if the nonresponse data have different characteristics than the original data then this assumption will not hold. A solution is to set some range for $s_2$, so that $(1 - \theta)s_1 \leq s_2 \leq (1 + \theta)s_1$, where $0 \leq \theta \leq 1$ and $\theta$ is set based on some prior inferences regarding the data. Given $s_2$, the effect size for the nonresponse data is given in (3).

---

[2] The only scenario not covered by the above is the "so-called" type III error scenario (Leventhal, and Huynh 1996), where the sample mean is in the opposite direction to the population mean.

$$d_2 = \frac{\bar{x}_2 - \mu_0}{s_2} \tag{3}$$

The sample mean for the nonresponse data is found by rearranging (3) to give (4).

$$\bar{x}_2 = d_2 s_2 + \mu_0 \tag{4}$$

This value can be used to find the sample mean for the combined response and nonresponse samples.

$$\bar{x}_c = \frac{n_1 \bar{x}_1 + n_2 \bar{x}_2}{n_1 + n_2} \tag{5}$$

The pooled standard deviation can be calculated using the meta-analysis formulation given in Higgins et al. (2019).

$$s_c = \sqrt{\frac{(n_1 - 1)s_1^2 + (n_2 - 1)s_2^2 + \frac{n_1 n_2}{n_1 + n_2}(\bar{x}_1^2 + \bar{x}_2^2 + 2\bar{x}_1 \bar{x}_2)}{n_1 + n_2 - 2}} \tag{6}$$

Consider the overall t test with the combined data for scenario 1. We wish to find the lowest $n_2$ for which $t \leq t^*$, and define some small quantity $\varepsilon$, such that $t + \varepsilon = t^*$, with $\varepsilon \geq 0$, so that $t = t^* - \varepsilon$.

$$t^* - \varepsilon = \frac{\bar{x}_c - \mu_0}{\frac{s_c}{\sqrt{n_1 + n_2}}} \tag{7}$$

Rearrange to put in terms of the total sample size.

$$\sqrt{n_1 + n_2} = \frac{s_c(t^* - \varepsilon)}{\bar{x}_c - \mu_0} \tag{8}$$

Rearrange to put in terms of $n_2$.

$$n_2 = \frac{s_c^2(t^* - \varepsilon)^2}{(\bar{x}_c - \mu_0)^2} - n_1 \tag{9}$$

The task is to find the smallest integer value of $n_2$ for which $\varepsilon \geq 0$. By making minor alterations, (9) can be used for scenarios (2)-(4). The four scenarios are summarized in Table 2.

Table 2: Scenarios for Single Sample t-test

| Scenario | Test Direction | Test Result | Find min{$n_2$} to make | $\varepsilon$ | Nonresponsed Bounded |
|---|---|---|---|---|---|
| 1 | Upper | Significant | Non-significant | $\varepsilon \geq 0$ | Upper |
| 2 | Upper | Non-significant | Significant | $\varepsilon < 0$ | Lower |
| 3 | Lower | Significant | Non-significant | $\varepsilon \leq 0$ | Lower |
| 4 | Lower | Non-significant | Significant | $\varepsilon > 0$ | Upper |

Each scenario has a direction of test (upper or lower), the result of the test on the response data, the opposite result, the range of $\varepsilon$ for which the minimum integer $n_2$ is being found, and how the nonresponse effect size is bounded[3]. Given that the t-value in (9) is dependent on $n_2$, giving a cross-dependency, $n_2$ cannot be calculated directly. A fixed-point optimization procedure for finding $n_2$ is given in the Appendix.

A similar process can be followed for two-sample independent sample tests. Sample sizes can be calculated for Student's t test (for equal variances) and Welch's t-test (for unequal variances) and a two-group measure, such as Glass's *g* or Cohen's *d* can be used to calculate effect sizes (Rosenthal & Rubin, 1982). If sample sizes are uneven, some constraints need to be placed on relative group sizes. For the sake of parsimony, full derivations are not included[4]. In addition, similar inference can be used for z tests of means and proportions.

---

[3] The exact bounds are not given here as there is a nonlinear dependence between sample size and effect size. For scenarios 1 and 3, there is an upper bound on effect size at which $n_2$ goes to infinity. For scenarios 2 and 4, there is a lower bound on effect size at which $n_2$ goes to infinity.
[4] These are available from the authors on request.

**Inference for Correlation Test**

Consider a situation, where a correlation is being tested for significance. The null hypothesis is $H_o: \rho = 0$, where $\rho$ is the population correlation. Standard alternate hypothesis are $H_a: \rho > 0$ (or $\rho < 0$) for a one-tailed test and $H_a: \rho \neq 0$ for a two tailed test. A population hypothesis is tested with a Pearson sample correlation coefficient $r$. The correlation $r$ is essentially an effect size (Cohen, 1988), with small ($0.1 \leq r < 0.3$), medium ($0.3 \leq r < 0.5$), and large ($r \geq 0.5$) effect sizes defined.

There are several different tests for the significance of correlations. The one most commonly used in meta-analysis involves transforming the correlation $r \in [-1,1]$ into a $z$ score using the inverse hyperbolic tangent transformation (Cox, 2008) and is given in (10).

$$zr_1 = \tanh^{-1}(r_1) = \frac{1}{2} \ln\left\{\frac{(1+r_1)}{(1-r_1)}\right\} \tag{10}$$

, where $r_1$ is the correlation coefficient for the response data. Now, this value is still essentially an effect size and does not depend on the sample size $n$. A standard error of $\sqrt{1/(n-3)}$ is defined by Fisher (1921), which can be used to give the z statistic in (11).

$$z_1 = \frac{zr_1}{SE(zr_1)} = \frac{zr_1}{\sqrt{1/(n-3)}} \tag{11}$$

Given that the z test is a simple two-way directional test, the four scenarios for finding the $n_2$ values needed to change a hypothesis test result are similar to the scenarios outlined for the one sample t-test. The only changes are that "z" replaces "t" for the test statistical and critical values and that the effect size defined for the nonresponse data is a correlation coefficient $r_2$, which can be transformed into a z score $zr_2$ using the same transformation as given in (10). The z-scores for the response and hypothesized nonresponse data can be combined (Field, 2001; Hedges & Vevea, 1998; Higgins et al., 2019) using (12).

$$zr_c = \frac{(n_1 - 3)zr_1 + (n_2 - 3)zr_2}{n_1 + n_2 - 6} \tag{12}$$

, which has the standard error given in (13).

$$SE(zr_c) = \sqrt{\frac{1}{n_1 + n_2 - 6}} \tag{13}$$

A similar process can be carried out as for the single sample t test. For scenario 1, we wish to find the lowest *n* for which $z \leq z^*$, where $z^*$ is the boundary value for significance and define some small quantity $\varepsilon$, such that $z + \varepsilon = z^*$, with $\varepsilon \geq 0$, so that $z = z^* - \varepsilon$.

$$z^* - \varepsilon = \frac{zr_c}{SE(zr_c)} = \frac{zr_c}{\sqrt{\frac{1}{n_1 + n_2 - 6}}} \tag{14}$$

This equation can be rearranged to give $n_2$.

$$n_2 = \left(\frac{z^* - \varepsilon}{zr_c}\right)^2 - n_1 + 6 \tag{15}$$

Now $n_2$ can be found in a similar manner to the single sample hypothesis test. The other three scenarios can be taken from Table 2 (but with *r* replacing *d* in the final column). A grid search optimization procedure for finding $n_2$ is given in the Appendix.

## Empirical Example

To assess the efficacy of the proposed WCRT, a simple survey was administered to a sample curated through Qualtrics. The goal of the survey was not to investigate any substantive empirical point, but to apply WCRT methods to assess robustness to participant nonresponse bias for a series of correlation tests. As the retailing scales employed by Szymanski & Henard (2001) have over 3,000 citations, and pose relatively simple questions, they were judged as liable to provide stable results, and unlikely to represent confounding factors due to their complexity.

**The Dataset**

  The survey includes five different multi-item measurement scales, each of which relates to some measure of customer satisfaction for a recent retail transaction. Each of the individual items is measured using a seven-point Likert scale. The full list of scales and items within these scales is given in Table 3. Overall, there are five different scales, consisting of 19 subitems. The first three scales deal with the actual shopping experience being evaluated, the fourth scale examines how this experience impacts behavioral intent, and the fifth is a general scale measuring retail/shopping enjoyment. Thus, the first three scales should be strongly correlated, while scale five may have some positive correlation with the other scales (someone who has positive views of retail shopping is more likely to select a positive shopping experience), but the level of correlation should be lower. Two of the items (item two on INTENT and item one on ENJOY) were negative direction items and were reversed.

Table 3: Survey Scale Information

| Information | Description |
|---|---|
| Name | Shopping Experience (EXP) |
| Prompt | Thinking about this retail shopping experience, please rate your overall feelings about the shopping experience. |
| Sub-items | unpleasant:pleasant<br>dislike very much:like very much<br>left me feeling bad:left me with a good feeling |
| Name | Satisfaction (SAT) |
| Prompt | My overall impression of this retail shopping experience is |
| Sub-items | Bad:Good<br>Unfavorable:Favorable<br>Unsatisfactory:Satisfactory<br>Negative:Positive<br>Dislike:Liked |
| Name | Positive Word of Mouth (PWOM) |
| Prompt | Thinking about your shopping experience, please rate your agreement with the following statements. |
| Sub-items | (All strongly disagree:strongly agree)<br>I would say positive things about this retailer.<br>I would recommend this retailer to people I know. |
|            | I would encourage relatives and friends to do business with this retailer.        |
| Name       | Behavioral Intentions (INTENT)                                                    |
| Prompt     | Thinking about your shopping experience, please rate your agreement with the following statements. |
| Sub-items  | (All strongly disagree:strongly agree) <br> I expect to be coming to this retailer for a long time. <br> I do not expect to visit this retailer in the future. <br> I expect my relationship with this retailer to be enduring. <br> It is likely that I will visit this retailer in the future. |
| Name       | Shopping Enjoyment (ENJOY)                                                        |
| Prompt     | Please rate your agreement with the following statements.                         |
| Sub-items  | (All strongly disagree:strongly agree) <br> I consider shopping a big hassle. <br> When traveling, I enjoy visiting new and interesting shops. <br> I enjoy browsing for things even if I cannot buy them yet. <br> I often visit shopping malls or markets just for something to do. |

The data were collected via a Qualtrics panel. There were $n = 415$ fully completed surveys out of a total of $n = 463$ surveys. In line with the focus on participant non-response bias, participant responses with missing participants were removed rather than imputed using a missing data technique.

**Exploratory Data Analysis**

As a preface to analyzing the correlation tests using WCRT, some analysis was performed on the consistency of the rating scale and on the correlations. To examine the consistency of the summated ratings scales, the Cronbach's alpha (Cronbach, 1951), was calculated for each of the summated rating scales. The values are EXP (0.96), SAT (0.99), PWM (0.96), INTENT (0.78), ENJOY (0.78). From past literature (e.g., Bland & Altman, 1997; Tavakol & Dennick, 2011), cut-offs for "good" to "excellent" values of alpha range from 0.7-0.95, so these values are in the correct range.

Figure 1: Multi-Item Scale Correlations

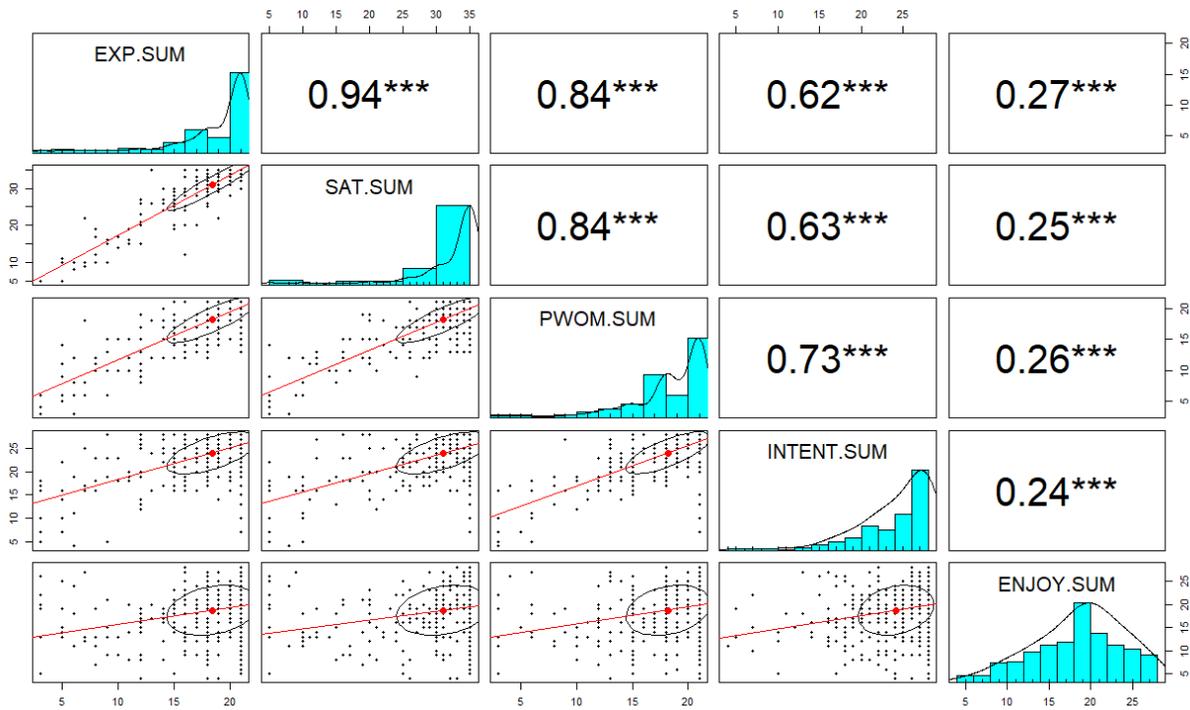

A summary matrix plot of the overall correlations between the values in the summated rating scales is given in Figure 1. Here, the diagonal values give histogram distributions of the summated values, the upper triangle of the matrix contains the correlations between the summated values (*** represents $p<0.001$ for a statistical test of correlation), and the lower triangle contains scatterplots, each overlaid with a linear regression best fit line and a confidence circle for the multivariate mean of the distribution.

**Wave Analysis**

A simple wave analysis was performed as follows. For this experiment, respondents were taken from a panel. As noted previously, there are $n = 415$ fully completed survey forms out of $n = 463$. For a panel, it is difficult to estimate the number of missing responses, but it is possible to estimate the percentage of missing participants given reported percentages for previous similar studies in the literature. For the purpose of this analysis, three scenarios were assumed, one with 50% response, one with 25% response, and one with 10% response.

The wave analysis approach described in Armstrong and Overton (1977), considers two different waves of responses, an early wave and a late wave, and then a "virtual" wave of nonresponses. While the responses to the survey were not split into waves, for the purpose of this illustrative example, it was assumed that the first 50% belong to the early wave and the second 50% belong to the late wave. The three response scenarios give the number of participant responses for the third wave as 415 (50% response), 1245 (25% response), and 3735 (10% response). Armstrong and Overton (1977) lay out three methods of calculating values for a third wave.

1. Assume that the nonresponses have the same mean as the second wave.
2. Assume that the nonresponses are at the same level as the responses at the end of the second wave.
3. Assume a linear interpolation through the nonresponse third wave.

For the measure of interest, let be the mean values for waves 1 and 2 respectively be $\bar{x}_1$ and $\bar{x}_2$. The number of item values in the three waves are denoted $n_1$, $n_2$, and $n_3$. Wave analysis aims to give a prediction for $\bar{x}_3$ in the nonresponse value. For case 1, $\bar{x}_3 = \bar{x}_2$. Cases 2 and 3 assume a linear relationship for the variable of interest over time. From Armstrong and Overton (1977)[5], for case 2, $\bar{x}_3$ is calculated as in (16).

$$\bar{x}_3 = \bar{x}_2 + (\bar{x}_2 - \bar{x}_1)\frac{n_2}{(n_1 + n_2)} \tag{16}$$

---

[5] To be consistent with the development of the WCRT method, the calculations are given using group means rather than upper and lower boundaries, but the calculations are equivalent.

Here, a straight line is drawn between the midpoint of group 1 and the midpoint of group 2. The line is extrapolated to the end of group 2. For the third scenario, the line is extrapolated to the middle of group 3, giving (17).

$$\bar{x}_3 = \bar{x}_2 + (\bar{x}_2 - \bar{x}_1)\frac{(n_2 + n_3)}{(n_1 + n_2)} \tag{17}$$

A wave analysis was performed for each combination of the correlations given in Figure 1 and the three different nonresponse scenarios. The first two wave analysis methods are independent of the number of nonresponses $n_3$, but the third is not, so can be calculated for all three nonresponse scenarios.

Table 4: Wave Analysis Results

| Correlation | $\bar{x}_1$ | $\bar{x}_2$ (M1) | M2: End wave 2 | M3: 415 (50%) | M3: 1245 (25%) | M3: 3735 (10%) |
|---|---|---|---|---|---|---|
| EXP, SAT | 0.928 | 0.955 | 0.969 | 0.997 | 1.000 | 1.000 |
| EXP, PWOM | 0.859 | 0.817 | 0.797 | 0.755 | 0.672 | 0.422 |
| EXP, INTENT | 0.709 | 0.517 | 0.421 | 0.230 | -0.154 | -1.000 |
| EXP, ENJOY | 0.313 | 0.225 | 0.181 | 0.093 | -0.084 | -0.612 |
| SAT, PWOM | 0.881 | 0.797 | 0.755 | 0.671 | 0.504 | 0.001 |
| SAT, INTENT | 0.736 | 0.498 | 0.380 | 0.142 | -0.333 | -1.000 |
| SAT, ENJOY | 0.287 | 0.204 | 0.163 | 0.080 | -0.086 | -0.585 |
| PWOM, INTENT | 0.815 | 0.625 | 0.531 | 0.342 | -0.036 | -1.000 |
| PWOM, ENJOY | 0.266 | 0.249 | 0.240 | 0.223 | 0.188 | 0.085 |
| INTENT, ENJOY | 0.264 | 0.210 | 0.183 | 0.129 | 0.021 | -0.304 |

The results of the wave analysis are given in Table 4. Results are given for each of the 10 possible correlations between the summated rating scales. The first two columns contain the values of the mean correlation values from waves one and two. The means for the second wave are taken as the M1 (method 1) estimate of the third wave. The next column contains the M2 estimate of the value at the end of the second wave and the subsequent columns contain the three M3 estimates for the three levels of participant response (50%, 25%, and 10%). For the moderate response scenarios, (50%, 25%), the correlations all stayed within bounds, but for the 10% scenario, several values need to be truncated at either -1 or 1. This shows the difficulty of a linear interpolation that extends well beyond the range of data. It is likely that as $n_3$ increases, any change in the dependent variable will lessen. However, the values for 10% response provide useful "extreme bounds", which can be utilized by the WCRT procedure.

**WCRT Procedure**

As previously shown in Figure 1, all multi-item scale correlations are strongly (p<0.001) significant, with correlations between scales related to the actual shopping experience (EXP, SAT, POW) being over 0.8, correlations between these scales and the future shopping intention (INTENT) scale being in the 0.6-0.7 range and the correlations between the general shopping enjoyment measure and the other scales being in the 0.2-0.3 range.

For each correlation, the WCRT procedure was calculated for effect sizes with increments of 0.01 ranging from -0.99 to the maximum effect size with a finite $n$ (-0.01) for α values from 0.01 to 0.1. Selected results are examined in Figures 2 and 3 in what we call "n-curves", which are similar to n-curves used to determine sample sizes (e.g., Trafimow 2018) or the probability of replication (Killeen, 2015) and the previously discussed p-curves for statistical power (Simonsohn et al., 2014).

For contrast, curves are given for the highest correlation (EXT and SAT) where $r_1 = 0.94$ and for the lowest correlation (INTENT and JOY) where $r_1 = 0.24$. For each of these correlations, curves are given for α = 0.05, though any value of α can be chosen. The x-axis contains the $r_2$ required to negate the significance of the significance test[6]. In the case of correlations, due to the asymptotic behavior of the significance test being a tradeoff between the overall effect size and $n$, only negative $r_2$ values give a finite $n$ and the graphs go off to infinity at approximately $r_2 = 0$.

---

[6] A similar n-curve could be drawn where the aim is to find $n$ to make a non-significant test significant.

As the relationship between the value of $r$ and $n$ is strongly exponential, it is difficult to plot $n$ versus $r$ on a linear scale, so a logarithmic scale is used for $n$. This makes it more difficult to read the values of $n$, but to make up for this, values of $n$ are explicitly given for negatives of the standard effect sizes defined by Cohen (1988), giving $r = -0.1$ (small), $r = -0.3$ (medium) and $r = -0.5$ (large) effect sizes, along with $r = -0.7$ and $r = -0.9$.

Looking at Figure 2, which is for an $\alpha = 0.05$ test for the pair of scales with the strongest correlation ($r_1 = 0.94$), for a small negative effect ($r_2 = -0.1$), $n = 5670$ would be required to negate significance, while for a large negative effect ($r_2 = -0.5$), $n = 1175$ would be required to negate significance. This would be very unlikely, given the large negative effect. Even an almost "complete reversal" of the correlation ($r_2 = -0.9$) would require $n = 454$ in order to negate significance.

Figure 2: n-Curve for EXP and SAT: α = 0.05.

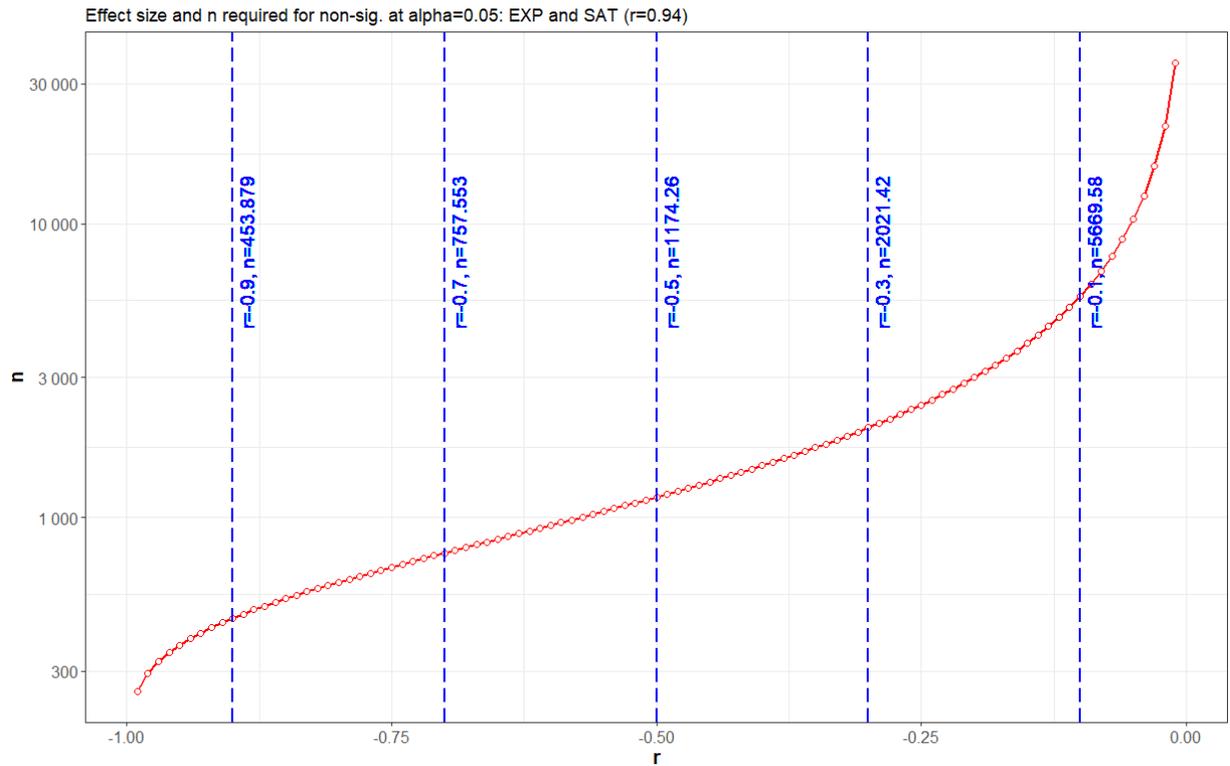

The graph in Figure 3 is for an α = 0.05 test for the lowest correlation of $r$ = 0.24 between INTENT and ENJOY with α = 0.05 and shows much lower values of $n$. For α = 0.05, for a small negative effect ($r_2$ = -0.1), $n$ = 427 would be required to negate significance, while for a large negative effect ($r_2$ = -0.5), $n$ = 103 would be required to negate significance. The extreme $r_2$ = -0.9 case would require $n$ = 43 to negate significance.

Figure 3: n-Curve for INTENT and ENJOY: α = 0.05.

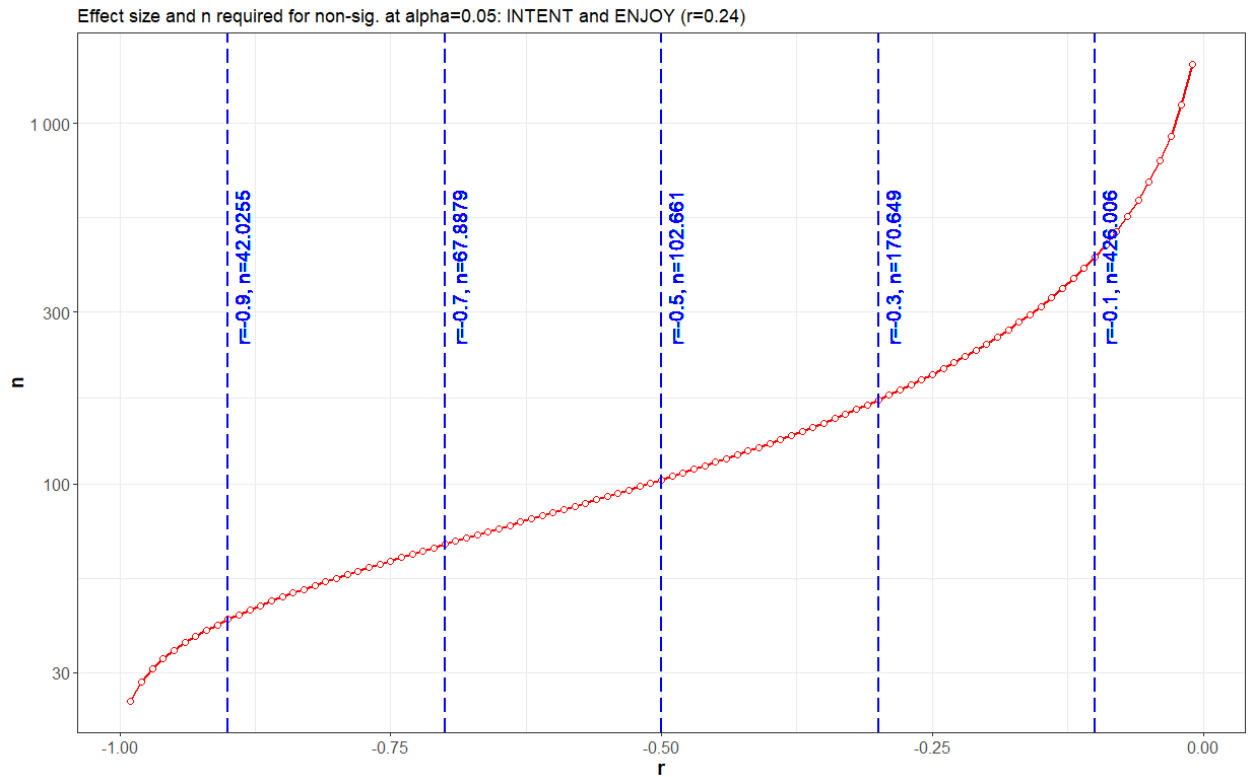

**Combining WCRT with Wave Analysis**

In any scenario where response times can be calculated, wave analysis can provide estimates of sample statistics for nonresponding participants, which can be converted to effect sizes. These effect sizes can be used to help choose a realistic range of effect sizes in the outlined WCRT procedure. Accordingly, we propose a method combining wave analysis results with WCRT to create a set of "warning" metrics for results that may be called in to question by possible nonresponse bias. An outline of the method is given below.

Assume a situation where a statistical test has been performed, with some level of Type I error $\alpha$ and there are two possible results; either $H_0$ is rejected in favor of $H_A$ or there is not enough evidence to reject $H_0$. The test will have some measure of effect size (e.g., Cohen's *d* for a two-sample test or the sample correlation $r$ for a correlation test). There is some number of nonresponses $n_3$ (for panel data, multiple scenarios can be given).

1. Calculate the three different wave analysis effect size values: M1: average of second wave, M2: end of second wave, M3: extrapolation to mean of third (nonresponse) wave.
2. For WCRT, calculate the effect size $r$ needed to reverse the statistical test given the number of nonresponses $n_3$. This is the inverse procedure of finding $n$ given an effect size, i.e., for a correlation $r$, if the calculation of $n$ from $r$ is defined as the function $f(r) = n$ then $f^{-1}(n) = r$.
3. Record if each of the three effect sizes found by wave analysis that will reverse the result of the statistical test. For example, for positive correlation $r$ that is statistically significant, if the correlation found by the WCRT is greater than the nonresponse correlation value predicted by wave analysis then the wave analysis correlation value will reverse the test.

The three wave analysis predictions give different levels of future extrapolations. For M1, where the predicted nonresponse mean is the mean of the second wave, unless a statistical result is close to a boundary, it is unlikely that a nonresponse mean effect size value equal to the value for the second wave will change the result of a statistical test. However, a linear extrapolation for M3 to the middle of the nonresponse wave for large nonresponse $n$ is liable to change a test result and the extrapolation is likely to be over-exaggerated, as it is unlikely that the trend from the 1st wave to the 2nd wave would continue linearly for a large nonresponse wave. Some damping is likely. However, the M3 scenario can provide a good "worst-case" scenario.

The combined method was applied to the previously discussed correlation example for all 10 correlations, two significance levels ($\alpha = 0.05, 0.01$), and the previously discussed participant nonresponse scenarios (nonresponse $n$ = 415, 1245, 3735). The results are given in

Tables 5 to 7, with each table containing one of the three nonresponse scenarios. Each table contains a row for each of the ten tested correlations. There are columns for the sample correlation value, the three wave analysis values, and the two WCRT values for the tested values of the Type I error α. As all correlations are significant and positive, the wave analysis results are flagged/counted as reversing the result of the statistical test if the correlations are less than the WCRT value.

Table 5: Combining Wave Analysis and Worst-case Resistance Testing for 50% Response (Nonresponse $n_3 = 415$)

| Correlation | r | $r_3$ (M1) | $r_3$ (M2) | $r_3$ (M3) | $Wr_3$ α=0.05 | $Wr_3$ α=0.01 |
|---|---|---|---|---|---|---|
| EXP, SAT | 0.94 | 0.955 | 0.969 | 0.997 | -0.924 | -0.917 |
| EXP, PWOM | 0.84 | 0.817 | 0.797 | 0.755 | -0.792 | -0.775 |
| EXP, INTENT | 0.62 | 0.517 | 0.421 | 0.230 | -0.526 | -0.495 |
| EXP, ENJOY | 0.27 | 0.225 | 0.181 | 0.093 | -0.139 | -0.097 |
| SAT, PWOM | 0.84 | 0.797 | 0.755 | 0.671 | -0.795 | -0.779 |
| SAT, INTENT | 0.63 | 0.498 | 0.380 | 0.142 | -0.536 | -0.505 |
| SAT, ENJOY | 0.25 | 0.204 | 0.163 | 0.080 | -0.114 | -0.072 |
| PWOM, INTENT | 0.73 | 0.625 | 0.531 | 0.342 | -0.662 | -0.637 |
| PWOM, ENJOY | 0.26 | 0.249 | 0.240 | 0.223 | -0.128 | -0.085 |
| INTENT, ENJOY | 0.24 | 0.210 | 0.183 | 0.129 | -0.104 | -0.061 |

In Table 5, no values are flagged and none of the wave analysis scenarios will reverse the result of the statistical test. In part, this is because all the test correlations are quite "strong". Even the correlations that include the ENJOY measure ($0.24 \leq r \leq 0.27$), while less than the other correlations, are strongly significant with a sample size of $n = 415$. As the nonresponse $n_3$ increases from 415 to 3735, the magnitude of the correlations found by the inverse WCRT procedure decrease. This is intuitive, as given that statistical significance is a function of both effect size and sample size, for a larger sample size, a smaller negative effect is needed to reverse the results of a statistical test.

Table 6: Combining Wave Analysis and Worst-Case Resistance Testing for 25% Response
(Nonresponse n3 = 1245)

| Correlation | r | r$_3$ (M1) | r$_3$ (M2) | r$_3$ (M3) | Wr$_3$ α=0.05 | Wr$_3$ α=0.01 |
|---|---|---|---|---|---|---|
| EXP, SAT | 0.94 | 0.955 | 0.969 | 1.000 | -0.475 | -0.428 |
| EXP, PWOM | 0.84 | 0.817 | 0.797 | 0.672 | -0.326 | -0.308 |
| EXP, INTENT | 0.62 | 0.517 | 0.421 | -0.154 | -0.173 | -0.154 |
| EXP, ENJOY | 0.27 | 0.225 | 0.181 | -0.084 | -0.028 (1) | -0.007 (1) |
| SAT, PWOM | 0.84 | 0.797 | 0.755 | 0.504 | -0.329 | -0.310 |
| SAT, INTENT | 0.63 | 0.498 | 0.380 | -0.333 | -0.178 (1) | -0.158 (1) |
| SAT, ENJOY | 0.25 | 0.204 | 0.163 | -0.086 | -0.019 (1) | 0.001 (1) |
| PWOM, INTENT | 0.73 | 0.625 | 0.531 | -0.036 | -0.241 | -0.221 |
| PWOM, ENJOY | 0.26 | 0.249 | 0.240 | 0.188 | -0.024 | -0.004 |
| INTENT, ENJOY | 0.24 | 0.210 | 0.183 | 0.021 | -0.016 | 0.005 |

In Table 6, the extrapolated r$_3$ for M3 goes outside of the testing "flip" boundaries defined by WCRT for three correlations, which increases to six correlations for the $n_3$ = 3735 results given in Table 7. This includes all the "enjoy" correlations except for the "PWOM, ENJOY" correlation, for which there is only a very slight linear trend. Despite negative linear trends, the "EXP, PWOM" and "SAT, PWOM" correlations are not flagged, as the correlations are high relative to the negative linear trends.

Table 7: Combining Wave Analysis and Worst-Case Resistance Testing for 10% Response
(Nonresponse $n_3$ = 3735)

| Correlation | r | r$_3$ (M1) | r$_3$ (M2) | r$_3$ (M3) | Wr$_3$ α=0.05 | Wr$_3$ α=0.01 |
|---|---|---|---|---|---|---|
| EXP, SAT | 0.94 | 0.955 | 0.969 | 1.000 | -0.158 | -0.148 |
| EXP, PWOM | 0.84 | 0.817 | 0.797 | 0.422 | -0.100 | -0.089 |
| EXP, INTENT | 0.62 | 0.517 | 0.421 | -1.000 | -0.046 (1) | -0.035 (1) |
| EXP, ENJOY | 0.27 | 0.225 | 0.181 | -0.612 | 0.003 (1) | 0.014 (1) |
| SAT, PWOM | 0.84 | 0.797 | 0.755 | 0.001 | -0.101 | -0.090 |
| SAT, INTENT | 0.63 | 0.498 | 0.380 | -1.000 | -0.047 (1) | -0.037 (1) |
| SAT, ENJOY | 0.25 | 0.204 | 0.163 | -0.585 | 0.006 (1) | 0.017 (1) |
| PWOM, INTENT | 0.73 | 0.625 | 0.531 | -1.000 | -0.069 (1) | -0.058 (1) |
| PWOM, ENJOY | 0.26 | 0.249 | 0.240 | 0.085 | 0.005 | 0.015 |
| INTENT, ENJOY | 0.24 | 0.210 | 0.183 | -0.304 | 0.007 (1) | 0.018 (1) |

**Discussion**

This study has presented a methodology and set of statistical tools for analyzing nonresponse bias situations. A methodology based on the file-drawer problem and worst-case resistance testing (WCRT) is given to help researchers quantify and understand the "robustness" of results with respect to nonresponse bias. Researchers can examine the number of non-responders to reverse the results of a statistical test for a range of feasible effect sizes for the nonresponse data. This relationship can be plotted using an "n-curve". The range of feasible effect sizes can be decided using evidence from past research, guidance on standard effect sizes, and the results of a wave analysis. Conversely, researchers can find the effect size needed to reverse the results of a statistical test for a given number of experimental nonresponses and then evaluate if these effect sizes are feasible using the guidance described above.

The basic WCRT methodology was developed in this paper as a method for analyzing robustness towards nonresponse bias. However, the methodology is more generally applicable to other scenarios. For any situation where there is a statistical test and some idea of possible "negative effect sizes", the WCRT methodology can be used to measure robustness. As noted in the introduction, there is a strong push to improve experimental rigor in the behavioral sciences and in marketing. An added urgency was added to this process by reports finding a low level of replicability in behavioral science studies (e.g., Open Science Collaboration 2015; Stanley et al. 2018) and by high-profile behavioral research scandals and retractions (e.g., Inman, et al. 2018; Stricker and Günther 2019). In addition to the focus on improving statistical rigor described earlier in the paper (e.g., JCR 2021; Harvey 2017; Schwab et al. 2011; Wasserstein and Lazar 2016), there has been a move towards requiring preregistration of experiments (Simmons et al. 2021), i.e., the process of researchers stating the experimental procedure and expected results and storing this information externally in a third-party repository, and to improved sharing and availability of research data (Towse et al., 2021). Including the preregistration information along with a paper submission ensures that the experiment is not altered in an ad-hoc manner to account for unexpected results.

The methods outlined in this paper can easily be incorporated into the behavioral science environment outlined above. Possible nonresponse bias should be noted, and procedures should be outlined for measuring bias. Even in a pure experimental setting, some type of nonresponse bias may be present; for example, for a student experiment, a certain number of students in a subject pool could be notified of a study, with only a few participating. When nonresponse bias is not an issue, WCRT can still be used to help examine the robustness of the results. Gelman and Loken (2013) noted that even with preregistration and no p-hacking, researchers can still bend the rules, for example, choosing the regression technique that gives the best results or choosing whether to use a main effect or interaction effect to justify a hypothesis. Given continued publication bias towards significant results (e.g., Franco et al. 2014; Harrison et al. 2017), there will always be an incentive to choose the research path to give the most significant results, in what statisticians sometimes call "the garden of forking paths". Rules to increase experimental rigor, such as preregistration, may prune some of these paths, but without being overly restrictive, cannot prevent researchers finding new paths. This is somewhat analogous to the situation of accountants finding new workarounds as rules on tax avoidance are strengthened.

In the context outlined above, WCRT could be utilized as a measure of robustness of results with respect to all possible experimental errors and biases. A range of possible effect sizes for the nonresponse bias could be derived and combined. Feasible nonresponse effect sizes could be derived for nonresponses using wave analysis or a similar method and generally by collating effect sizes the past literature in the area or through a meta-analytic p-curve analysis (Simonsohn et al. 2014). In time, a set of "n" thresholds could be developed to flag results with insufficient robustness to the factors outlined above.

**Limitations and Future Research**

This paper develops WCRT methods for correlations and single-sample hypothesis tests[7]. To be widely utilized, WCRT methods would need to be developed for a wider range of statistical tests, such as regression, ANOVA and SEM, as these methods are the most widely used methods in behavioral research. This is a similar scenario to effect size and power calculations, where over time, methods have been developed for a wide range of statistical tests. For the methods to be widely used, it would be important to package them together into a single cohesive software package, in a similar manner to G*Power (Faul 2007), which has become the de-facto standard software package for power analysis.

In the modern internet-mediated environment, more surveys are being conducted using online panels designed to represent certain population characteristics and through co-working/online hiring platforms, such as the Amazon Mechanical Turk (Kees et al. 2017). Determining nonresponse in online environments is difficult, as the survey platform recruitment procedure may be opaque. What exactly constitutes nonresponse in a panel or online working platform? If a set of respondents are notified about an opportunity, then the number of nonresponses can be calculated only if the number notified is reported by the platform. In a co-working platform where respondents search through lists of opportunities, calculating nonresponse may be more difficult. If views of an opportunity are recorded (e.g., through a scroll-down list), then some measure of nonresponse of "aware" respondents can be calculated, but determining how to set a threshold for awareness would be difficult. There has been some initial work on analyzing nonresponse for the Mechanical Turk for longitudinal studies (Daly and Nataraajan 2015) and several studies have tried to quantify possible nonresponse bias for online platforms (Boas et al. 2020; Paolacci et al. 2010). However, there is strong scope for a systematic analysis of nonresponse for online surveys. Such analysis could include work from both information systems and experimental standpoints and include aspects such as data reporting, human-computer interaction, and nonresponse behavior.

---

[7] Two sample hypothesis tests have also been developed and material is available from the authors on request.

The wave analysis method utilized in this paper is a simple linear extrapolation method. Linear extrapolation may not be reliable outside of the range of the data. It is likely that significant linear trends would probably "damp" outside of the range of the data, particularly in situations where there are many non-respondents. This is a reason why damped trend forecasting methods that give conservative forecasts are often successful (e.g., Armstrong et al. 2015; Gardner 2015). In the use of wave analysis in the experimental section, this lack of conservatism is an advantage, as linear extrapolation is used to create worst case bounds for correlations. However, given the advances in forecasting over the 40 plus years since the introduction of wave analysis (e.g., Makridakis et al. 2020), there is scope to bring new methodology to bear on wave analysis and develop methods to improve forecasts of nonresponse bias.

# APPENDIX A: Optimization Algorithms

**Optimization Algorithm for Single Sample t-Test Inference**

Equation (9) in the main paper cannot be solved outright as both $s_c$ and $t^*$ are dependent on $n_2$, creating cross-dependencies. However, the equation can be solved using a simple fixed-point algorithm.

1) Utilize an initial starting value of $n_2 = n_1$ and call this *nOpt*.

2) Calculate $\bar{x}_2, \bar{x}_c$, and $s_c$ using *nOpt*.

3) Calculate $n_2$ from equation (9) and store this in variable *nCalc*.

4) Recalculate *nOpt* as (*nOpt*+*nCalc*)/2.

5) Repeat steps 2-4 until |*nOpt-nCalc*|<$\delta$, where $\delta$ is some pre-set convergence criterion.

In practice, the values of *nOpt* and *nCalc* always converge so that |*nOpt-nCalc*|<$\delta$.

**Optimization Algorithm for Correlation Inference**

The fixed-point method used for the previous tests did not converge for the correlation test, due to Equation (16) in the main paper having both a negative and positive root. Thus, a divide and conquer optimization method was employed. It takes advantage of the fact that given a candidate value of $zr_c$, the value of $n_2$ can be calculated by rearranging Equation (12) in the main paper as follows:

$$zr_c(n_1 + n_2 - 6) = (n_1 - 3)zr_1 + (n_2 - 3)zr_2 \tag{A-18}$$

Collecting $n_2$ terms gives (A-2).

$$n_2(zr_c - zr_2) = (n_1 - 3)zr_1 - 3zr_2 - zr_c(n_1 - 6) \tag{A-19}$$

Rearranging in terms of $n_2$ gives (A-3).

$$n_2 = \frac{(n_1 - 3)zr_1 - 3zr_2 - zr_c(n_1 - 6)}{(zr_c - zr_2)} \tag{A-20}$$

1. The algorithm works by exploring the possible values of $zr_c$, calculating $z$ and then constraining z towards $z^* \pm \varepsilon$. Calculate $zr_1$ from $r_1$. Calculate $zr_2$ from $r_2$. For a nonresponse effect size $r_2$, the steps are as follows:

2. From (12) in the main paper, $zr_c$ is a linear combination of $zr_1$ and $zr_2$, so lies between these two values. For cases 1 and 4, set LB = $zr_1$ and UB = $zr_2$; for cases 2 and 3 set LB = $zr_2$ and UB = $zr_1$.

3. Set $zr_c = \dfrac{LB + UB}{2}$ and then use this value of $zr_c$ to calculate $n_2$, using (A-3).

4. Calculate the value of $z = zr_c \sqrt{n_1 + n_2 - 6}$.

5. Now, if $z < z^*$, set $LB = zr_c$ else set $UB = zr_c$.

6. If $|UB - LB| < \delta$, where δ is a convergence criterion then exit. Otherwise go to step 3.

If the value of $z^*$ is not in the range of the initial [LB,UB] this indicates that there is no possible $n_2$ for the selected $r_2$ that can give a z that reaches the critical value. Typically, for case 1 and 4, when increasing effect size, this occurs when $r_2$ is just above 0, i.e., where the trade-off of the low effect size against high $n_2$ reaches a possible equilibrium.

**Open Practices Statement:** All code and data used in this paper have made available at https://github.com/MDSOPT/WCRT.